\documentclass[12pt]{article}
\usepackage{epsf}

\title{ On-line learning in a discrete state space
      }

\author{W. Kinzel and R. Urbanczik  \\
        Institut f\"ur theoretische Physik\\
        Universit\"at W\"urzburg \\
        Am Hubland\\
        D-97074 W\"urzburg \\
        Germany
       }

\newcommand{\sign}{ \mbox{\rm sign} }

\begin{document}

\baselineskip 1.8\baselineskip

\maketitle

\begin{abstract}
  \setlength{\parindent}{0.0cm}
  On-line learning of a rule given by an N-dimensional Ising perceptron,
  is considered for 
  the case when the student is constrained to take values in a discrete state
  space of size $L^N$. For $L=2$ no on-line algorithm can achieve a 
  finite overlap with the teacher in the thermodynamic limit.  
  However, if $L$ is on the order of $\sqrt{N}$, Hebbian learning does
  achieve a finite overlap.
  \end{abstract} 

Artificial neural networks are usually trained by a set of examples
\cite{Her91}. After the training phase such a network (="student") has
achieved some knowledge about the rule (="teacher") which has
generated the examples. The difference between the outputs of the
student and the teacher for a random input vector defines the
generalization error.

There are two basic kinds of training algorithms: 1. In batch
mode the complete set of examples is stored and iteratively used to
change the synaptic weights of the student network. 2. In on-line
mode each example is used only once. At each training step a new
example is presented and the synaptic weights are changed according
to some algorithm. 

The analysis of on-line algorithms using methods of statistical
mechanics \cite{Bie93,Kim96,Kin92,Kin90,Saa95} has shown that this is a 
powerful and versatile
approach to learning problems. To our knowledge, however, only
continuous couplings have so far been considered. But for hardware
implementations it would be extremely useful to design algorithms
which work in a discrete space of synaptic weights. It is not known
whether on-line algorithms work at all for weights which have a
limited number L of possible values. Here we show for a simple case
that generalization is only possible if L is of the order of
$\sqrt{N}$, where $N$ is the size of the network.

We consider the perhaps simplest learning scenario in which the
teacher is a perceptron with $N$ binary
couplings $B_i \in \{-1,1\}$. In on-line learning, the student
perceptron with weight vector $J$ receives at each time step an
N-dimensional input $\xi$ and the classification bit $\sigma_B(\xi)
\in \{-1,1\}$ provided by the teacher $B$. The task is to find a
mapping, $J' = f(J,\xi, \sigma_B(\xi))$ which updates the student $J$,
our current approximation of $B$, based on this information. Of
course, $J'$ should be an improved approximation. Under very general
conditions, we show in this Letter that no such mapping exists if $J$
and $J'$ are confined to lie, as the teacher is, in the set 
$\{-1,1\}^N$.  In a second step, we consider Hebbian learning in a
discretized state space of size $L^N$, and determine the
generalization behavior as function of $\lambda=L/\sqrt{N}$.

The classification of $\xi$ is given by $\sigma_B(\xi) = \sign(B^T \xi)$.
Hence the quality of the approximation provided by a student $J$ can be
defined via the overlap $R = N^{-1}B^T J$ with the teacher. Since the students
have binary components, it is convenient to have the update rule 
$f$ specify at which
sites the sign should be flipped to obtain the updated weight vector $J'$. 
So $J'_i = J_i f_i(J,\xi, \sigma_B(\xi))$ and the $f_i$ take values in
$\{-1,1\}$.  The update rule will be useful
if it improves on our current state, that is if
\begin{equation}
B^T J' = \sum_{i=1}^N B_i J_i f(J,\xi, \sigma_B(\xi)) > B^T J \;.
\end{equation}
Of course $f$ cannot have any built in knowledge about the teacher
but must infer information about $B$ from the current pattern.
Formally this can be enforced by requiring that $f$ be useful not
just for the single teacher $B$ but on average, for teachers which
have the same overlap as $B$ with $J$. Denoting by 
$\langle \ldots \rangle_{B|B^T J = NR}$ the average over the uniform
distribution on the set of teachers which have overlap $R$ with $J$, a
useful $f$ must thus fulfill: 
\begin{equation}
\left\langle\sum_{i=1}^N B_i J_i f_i(J,\xi, \sigma_B(\xi))
\right\rangle_{B|B^T J = NR}
  > NR \;. \label{cond}
\end{equation}
By a gauge transformation the LHS may be written as 
$\langle\sum_{i=1}^N B_i f_i(J,\xi, \sigma_B(\xi^*))\;
 \rangle_{B|\sum_i B_i = NR}$ where $\xi^*$ is given by
$\xi^*_i = J_i\xi_i$. Using that for the Heaviside step function
$\theta$, 
$1 = \theta(\sigma_B(\xi^*)) + \theta(-\sigma_B(\xi^*))$, we may
rewrite (\ref{cond}) as
\begin{equation}
\sum_{\sigma \in \{-1,1\}}\sum_{i=1}^N  f_i(J,\xi, \sigma)
\left\langle B_i \theta(\sigma B^T \xi^*)\right\rangle_{B|\sum_i B_i = NR}
  > NR \;. \label{condb}
\end{equation}
Under mild conditions on $\xi$, one finds that 
\begin{equation}
\langle B_i \theta(\sigma B^T \xi^*)\;\rangle_{B|\sum_i B_i = NR} \ge 0
\label{conda}
\end{equation}
for any positive $R$ in the limit of large $N$. Consequently the LHS of
(\ref{condb}) is maximized by choosing $f_i(J,\xi, \sigma)=1$, 
and the best we can do is to keep the weight vector $J$ fixed.

There are some special cases, where (\ref{conda}) is not true. If just a 
single component of $\xi$ is nonzero, then $\sigma_B(\xi)$ will of course
give us the corresponding component of $B$ and one can achieve
$R=1$ by asking $N$ such questions. But it is hard so see how such a 
strategy might be extended to the case of a noisy teacher.

For more generic patterns,
however, the $\xi_i$ will be of similar magnitude. Further,
$\xi$ will only have
a small overlap with $J$, that is $m = \sum_i \xi_i J_i/|\xi|$ will be of
order 1. Then for large N, and consequently small $\xi_i/|\xi|$,
the  LHS of (\ref{conda}) may be evaluated 
using the central limit theorem and yields 
\begin{equation}
\langle B_i \theta(\sigma B^T \xi^*)\;\rangle_{B|\sum_i B_i = NR} = 
 R H\left(-\sigma m \frac{R}{\sqrt{1-R^2}} -
 \frac{\sqrt{1-R^2}}{R}\frac{\sigma\xi_i J_i}{|\xi|}\right), \label{largeN}
\end{equation}
which is positive. So if the components of $\xi$ are picked independently
from distributions having bounded ratios of their variances,
the fraction of inputs for which
(\ref{conda}) is violated decreases exponentially with $N$.

An even stronger statement can be made for binary inputs, 
$\xi_i \in \{-1,1\}$. Then the large $N$ expansion yielding
(\ref{largeN}) can only be wrong, if the input is  correlated with
the student $(|m| \gg 1)$. But for this case  (\ref{conda}) may be
verified by evaluating its LHS with the saddlepoint method. Consequently,
for binary inputs, on-line learning is impossible even if queries 
\cite{Kin90} are allowed.

As it is possible to learn on-line with continuous couplings, the
question arises what the numerical depth of the couplings must be, for
on-line learning to succeed. We thus consider a situation where the
$J_i$ are constrained to lie in the set $\{1,2,\ldots,L\}$, still with
a binary teacher.  A weight vector $J$ is then taken to represent an
estimate $\tilde{B}$ of $B$ via $\tilde{B}_i = \sign(J_i - L/2)$. For
randomly chosen binary inputs, Hebbian learning may be applied to J by
truncating to the allowed range of values:
\begin{equation}
 J_i' = \left\{
        \begin{array}{ll}
         J_i + \xi_i \sigma_B(\xi)&\mbox{if\ } 
         J_i + \xi_i \sigma_B(\xi) \in \{1,\ldots,L\} \\
         J_i &\mbox{else.}
        \end{array}
        \right. \label{hebb}
\end{equation}
The increments $\xi_i \sigma_B(\xi)$ are not independent over the
sites $i$ but their covariances do decay as $1/N$. So for large $N$
the sites will approximately decouple, and we are left with a biased
random walk on each site. The bias is given by 
\begin{equation} 
< \xi_i \sigma_B(\xi)> = B_i \sqrt{\frac{2}{\pi N}}
\end{equation}
where $<...>$ is an average over random vectors $\xi$.

Let $p_l(t)$ denote the probability that
$J_1 = l$ after $t$ iterations of (\ref{hebb}) and assume that 
$B_1 = 1$. Then
\begin{eqnarray}
p_1(t+1) &=& r p_1(t) + r p_2(t) \nonumber\\
p_l(t+1) &=& g p_{l-1}(t) + r p_{l+1}(t)\,, \mbox{\ \ } l = 2,\ldots,L-1
             \nonumber \\
p_L(t+1) &=& g p_{L-1}(t) + g p_L(t) \;, \label{walk}
\end{eqnarray}
where $r+g=1$ and $g = 1/2 +1/\sqrt{2 \pi N}$ for large $N$. The stationary
solution $p^{\rm s}$  of (\ref{walk}) is 
$p^{\rm s}_l \propto  (g/r)^l$. Thus for large $N$
the asymptotic overlap $R^{\,\rm s}$ between the estimate $\tilde{B}$ and 
the teacher will approach zero if $L$ is fixed. For $L= \lambda \sqrt{N}$,
however, one finds
\begin{equation} 
 R^{\,\rm s} = 1 - \frac{2}{1+e^{\sqrt{8/\pi}\;\lambda}}\;.
\end{equation}

The time needed to approach the stationary distribution will scale linearly
with $N$ for fixed $\lambda$. So let $R(\alpha)$ be the overlap after
$\alpha N$ steps, assuming that initially $J_i = L/2$. The time evolution
of $R$ may then be calculated using the
explicit formulas for the powers of the transition matrix of the 
random walk (\ref{walk}) given in \cite{Fel51}. One finds:
\begin{equation}
R(\alpha) = R^{\,\rm s} - 4\sqrt{2/\pi} e^{-\alpha/\pi}
            \sum_{k=0}^{\infty} 
              e^{-\frac{\pi^2\alpha}{2\lambda^2} (2k+1)^2}
              \frac{\lambda}{\frac{2}{\pi}\lambda^2 + \pi^2 (2k+1)^2}\;. 
\label{ralpha}
\end{equation}
The resulting dependence of the overlap on $\lambda$  
(for fixed $\alpha$) is nonmonotonic as shown in Figure 1. 
For large $\alpha$ the sum in the above expression is dominated by
the first term and $R$ decays exponentially; this gives the relaxation time 
\begin{equation}
\tau = \frac{2\lambda^2\pi}{2\lambda^2 + \pi^3}.
\end{equation} 
To find the behavior for $L \gg \sqrt{N}$, we need
to take the limit $\lambda \rightarrow \infty$ in (\ref{ralpha}),
that is, replace the sum over $k$ by an integral. This yields
\begin{equation}
R(\alpha) = 1 - 2 H(\sqrt{2\alpha/\pi})\;,
\end{equation}
the result found in \cite{Broe93} for the case, where one applies Hebb's rule
to continuous couplings and clips in the end.

We have considered only simple Hebbian learning here. However, 
since $\alpha N$ examples will be needed to achieve good generalization,
we believe that one cannot improve on the scaling, $L=\lambda \sqrt{N}$, 
by using a different algorithm.

One of the authors (W.K.) would like to thank Ido Kanter for useful
discussions. The work of R.U. was supported by the {\em Deutsche 
Forschungsgemeinschaft} (DFG).     

\bibliographystyle{plain}
\bibliography{../tex/neural}

\pagebreak
\begin{figure}
\begin{tabular}[b]{cc}
          \large$R$&\parbox{10.2cm}{\epsfbox{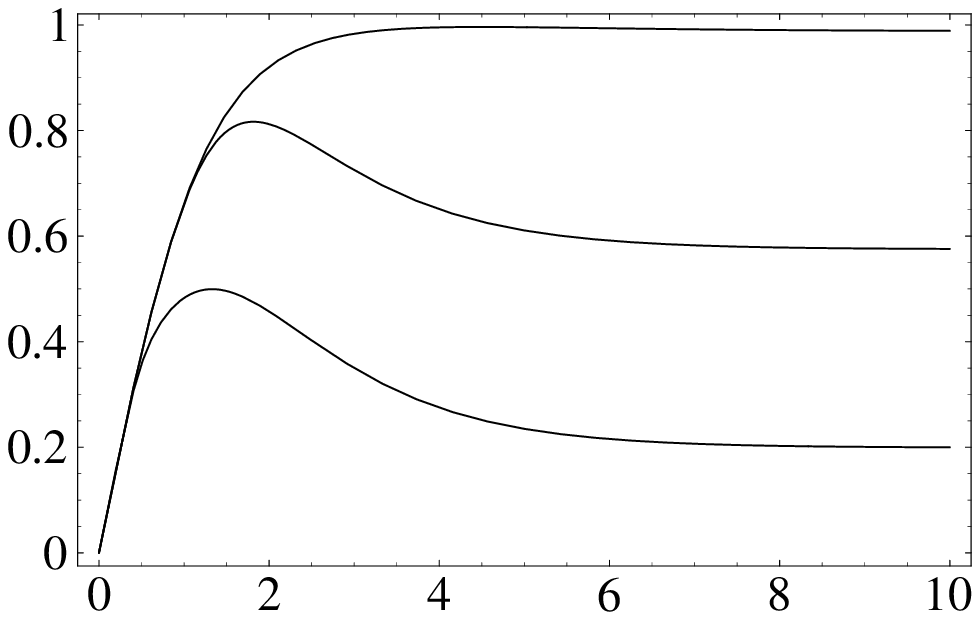}}\\
        &\Large \hspace{1cm}$\lambda$                 
\end{tabular}
\parbox{11.5cm}{
  \caption{Overlap $R$ achieved by Hebbian learning using $L$ different
           weight values per coupling, $L= \lambda \sqrt{N}$. The
           curves are, from top to bottom, for 
           $\alpha = 10$, $\alpha = 1$ and $\alpha = 0.1$
           }
}
\end{figure}

\end{document}